\documentclass[10pt, conference]{IEEEtran}
\IEEEoverridecommandlockouts
\usepackage[numbers, sort&compress]{natbib}
\usepackage{amsmath, amssymb, amsthm, amsfonts}
\usepackage{mathtools}
\usepackage{booktabs}
\usepackage{graphicx}
\usepackage{array}
\usepackage{float}
\usepackage{algorithm}
\usepackage{algpseudocode}
\usepackage{subcaption}
\usepackage{dblfloatfix}
\usepackage{titlesec}
\usepackage{multirow}
\usepackage{booktabs}

\titlespacing*{\section} {0pt}{1.5ex}{1ex}
\titlespacing*{\subsection} {0pt}{1ex}{1ex}

\AtBeginDocument{%
    \setlength\intextsep{3pt}%
    \setlength\textfloatsep{5pt}}
\begin{document}

\title{Intelligent Edge Resource Provisioning for\\ Scalable Digital Twins of Autonomous Vehicles} 

\author{\IEEEauthorblockN{
Mohammad Sajid Shahriar\IEEEauthorrefmark{1},
Suresh Subramaniam\IEEEauthorrefmark{2},
Motoharu Matsuura\IEEEauthorrefmark{3}, \\
Hiroshi Hasegawa\IEEEauthorrefmark{4}, and Shih-Chun Lin\IEEEauthorrefmark{1}
}\vspace{0.5em}
\IEEEauthorblockA{\IEEEauthorrefmark{1} Intelligent Wireless Networking (iWN) Laboratory, Department of Electrical and Computer Engineering, \\
North Carolina State University, NC, USA. Email: \texttt{\{mshahri,  slin23\}@ncsu.edu}\\
\IEEEauthorrefmark{2} George Washington University, WA, USA. Email: \texttt{suresh@gwu.edu}\\
\IEEEauthorrefmark{3} University of Electro-Communications, Tokyo, Japan. Email: \texttt{m.matsuura@uec.ac.jp}\\
\IEEEauthorrefmark{4} Nagoya University, Nagoya, Japan. Email: \texttt{hasegawa@nuee.nagoya-u.ac.jp}}

}

\maketitle
\begin{abstract}
The next generation networks offers significant potential to advance Intelligent Transportation Systems (ITS), particularly through the integration of Digital Twins (DTs). However, ensuring the uninterrupted operation of DTs through efficient computing resource management remains an open challenge. This paper introduces a distributed computing architecture that integrates DTs and Mobile Edge Computing (MEC) within a software-defined vehicular networking framework to enable intelligent, low-latency transportation services. A network aware scalable collaborative task provisioning algorithm is developed to train an autonomous agent, which is evaluated using a realistic connected autonomous vehicle (CAV) traffic simulation. The proposed framework significantly enhances the robustness and scalability of DT operations by reducing synchronization errors to as low as 5\% while achieving up to 99.5\% utilization of edge computing resources.
\end{abstract}


\setlength{\abovecaptionskip}{3px}
\setlength{\belowcaptionskip}{-5px}

\section{Introduction}

The emerging 6G networks is expected to greatly improve Intelligent Transportation Systems (ITS) by offering faster data exchange, lower delays, and more stable connections. One of the key technologies supporting this progress is the Digital Twin (DT)—a real-time, virtual model of a physical system that continuously collects and processes data from the real world \cite{DigitalTwin_sensing}. DTs enable the analysis and prediction of system behavior without interfering with the physical environment. In complex and rapidly evolving scenarios such as the Internet of Vehicles (IoV), they offer an efficient and cost-effective means of monitoring, testing, and optimizing system performance \cite{DT_roundabout}. When integrated with data-intensive and large-scale sensor sources from autonomous vehicles, such as cameras, LiDAR, and radar \cite{khan2025vehicle}—digital twins provide a powerful platform for developing and validating artificial intelligence (AI) models, evaluating traffic safety strategies, and facilitating cooperative decision-making between vehicles and infrastructure, all without impacting real-world traffic operations \cite{shahriar2023drl}. However, supporting such data-driven applications requires ultra-low latency and fast response times.

To meet these demands, Mobile Edge Computing (MEC) brings computational resources closer to vehicles and roadside units, minimizing delays caused by communication with distant cloud servers. While this reduces latency and improves responsiveness, managing the increasing complexity and scale of such edge-based networking systems remains a significant challenge. Software-Defined Networking (SDN) addresses this by offering a more adaptable and centralized approach to network management. By decoupling the data forwarding layer from the control layer, SDN leverages intelligent controllers or computing devices that dynamically manage network behavior and resource allocation. As the computing devices continue to become more powerful, a key research challenge lies in how to integrate and coordinate these technologies—digital twins, MEC, and SDN—within a unified computing framework. Addressing this challenge would not only streamline the design and operation of such complex systems but also promote scalable and efficient use of connected infrastructure in future intelligent transportation environments.

This paper addresses the above challenge by proposing a distributed computing architecture that integrates edge computing capabilities within a software-defined vehicular networking framework. This integrated infrastructure combines communication and computation to enable a novel and dynamic resource provisioning method for data-driven services with ultra-low latency. The main contributions of this work are summarized as follows:

\begin{enumerate}
\item We propose a two-tier architecture for DT, where the data-intensive processing is carried out in a distributed manner, while ensuring efficient control of data transmission within the same computing domain. 
\item We model the synchronization of DTs with their connected autonomous vehicle (CAV) counterparts, as well as collaborative task computation using an edge network, in order to achieve efficient DT synchronization.
\item We develop a scalable framework including a DRL algorithm to train an autonomous agent, which utilizes the collaborative task computation model to efficiently balance the control and computational capabilities of edge infrastructures.
\item We evaluate the performance of the trained agent using a CAV traffic simulation developed using a real-world map and empirical traffic volume data.  
\end{enumerate}

\section{Related Works}

Several research efforts have explored different aspects of communication, computation integration, and DT technologies in vehicular and edge computing environments. The authors of \cite{zhang2024two} investigate the optimization of data-intensive services in communication and computation integration networks. They propose a two-stage resource scheduling scheme that jointly performs resource orchestration and transmission scheduling in an online manner. It is designed to adapt to network dynamics across various dimensions, including spatial variations in computing resources and temporal fluctuations in time slot availability. However, the greedy approach commonly used in these studies for computing resource scheduling do not address the synchronization challenges inherent to digital twin systems.

\begin{figure}[t]
  \centering
    \includegraphics[width=3.4in]{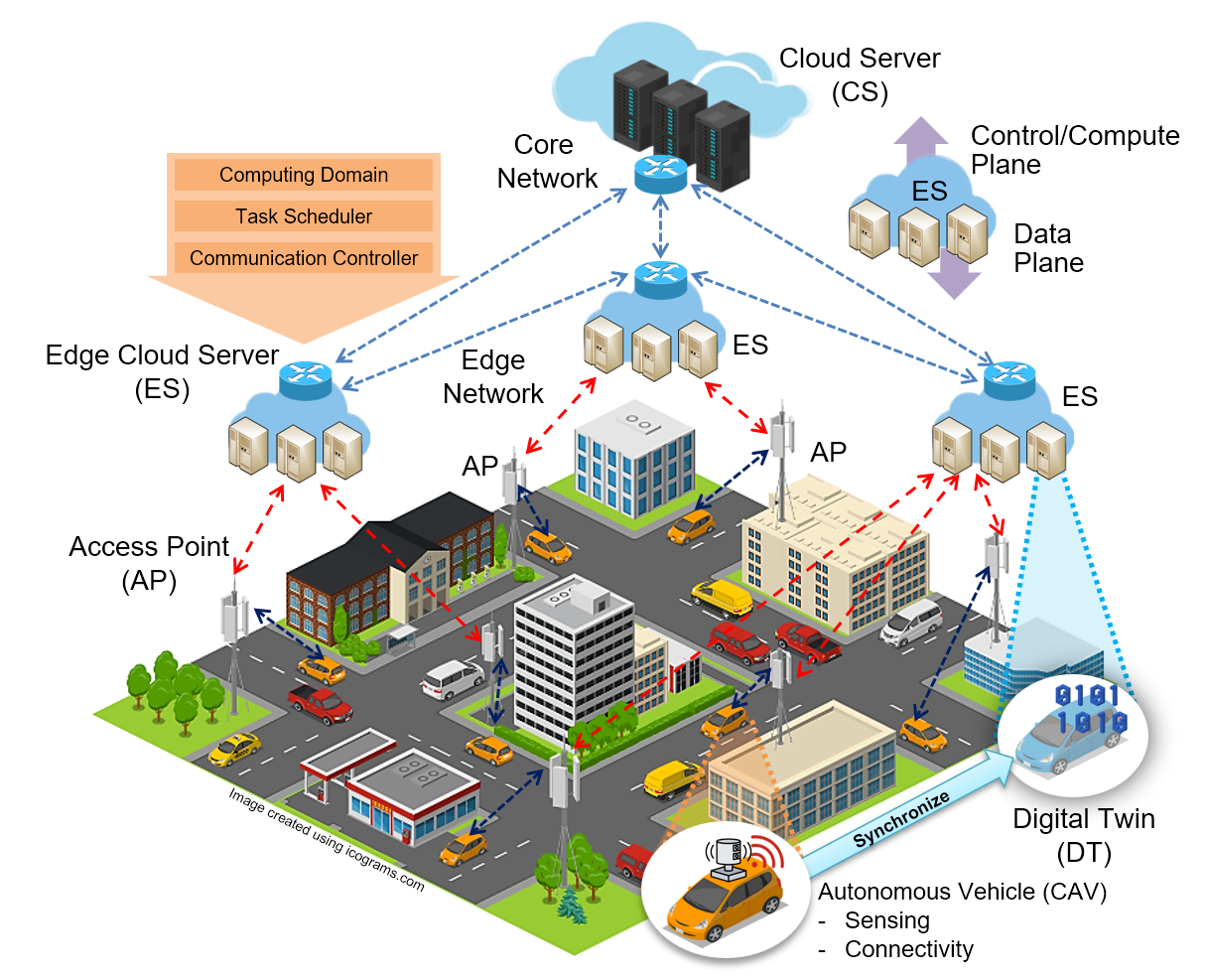} 
  \caption{System model of edge network enabled DTs of CAV.} 
  \label{fig:system model} 
\end{figure}

The work in \cite{khan2025vehicle} introduces a framework for DT-assisted cloud-edge collaborative IoVs tailored for intelligent transportation systems. This study considers the influence of vehicular sensing on DT synchronization accuracy and data redundancy, emphasizing that the performance of DTs is significantly affected by their deployment. Furthermore, the authors develop a five-dimensional system Quality of Service (QoS) model and formulate an optimization problem that jointly optimizes vehicular sensing and virtual DT VDT deployment to maximize system QoS. Nevertheless, this work does not delve into resource allocation mechanisms for edge infrastructures, which are critical in practical deployments.

Additionally, reinforcement learning (RL) algorithms have been leveraged for complex joint optimization problems. For example, \cite{pervej2020eco} demonstrates the effectiveness of RL in addressing hard combinatorial optimization tasks. Their findings suggest that a distributed learning strategy operating over a centralized action space yields superior performance compared to conventional multi-agent RL (MARL) models. Despite these promising results, scalability remains a significant limitation, and there is a pressing need for frameworks that integrate scalability with the problem-solving capabilities of RL-based methods.

\section{System Design}
\subsection{Scalable Communication-Computation Framework}
To facilitate DTs of CAVs, we propose resource provisioning considering a network framework that integrates communication and computation through distributed computing as defined in \cite{lin2022sdvec}. We consider a set of edge cloud servers (ESs) that are linked through an edge networks depicted in Fig. \ref{fig:system model}. The ESs manage $E$ geographically distributed computing domains, represented by $\mathcal{E} = \{1,2,\ldots,E\}$. Each $e \in \mathcal{E}$ also controls $R$ remote access points (APs), which are represented as $\mathcal{R}_e = \{1,2,\ldots,R\}$.
Each access point $r \in \mathcal{R}_e$ supports $V$ number of CAVs for communication, represented as $\mathcal{V}_r = \{1,2,\ldots,V\}$. A computing cluster is thus defined as an edge computing domain comprising an ES together with it's associated APs and the set of CAVs they serve. 

\subsection{Two-Tier Architecture for DT of CAV}
\begin{figure}[t]
  \centering
    \includegraphics[width=3.4in]{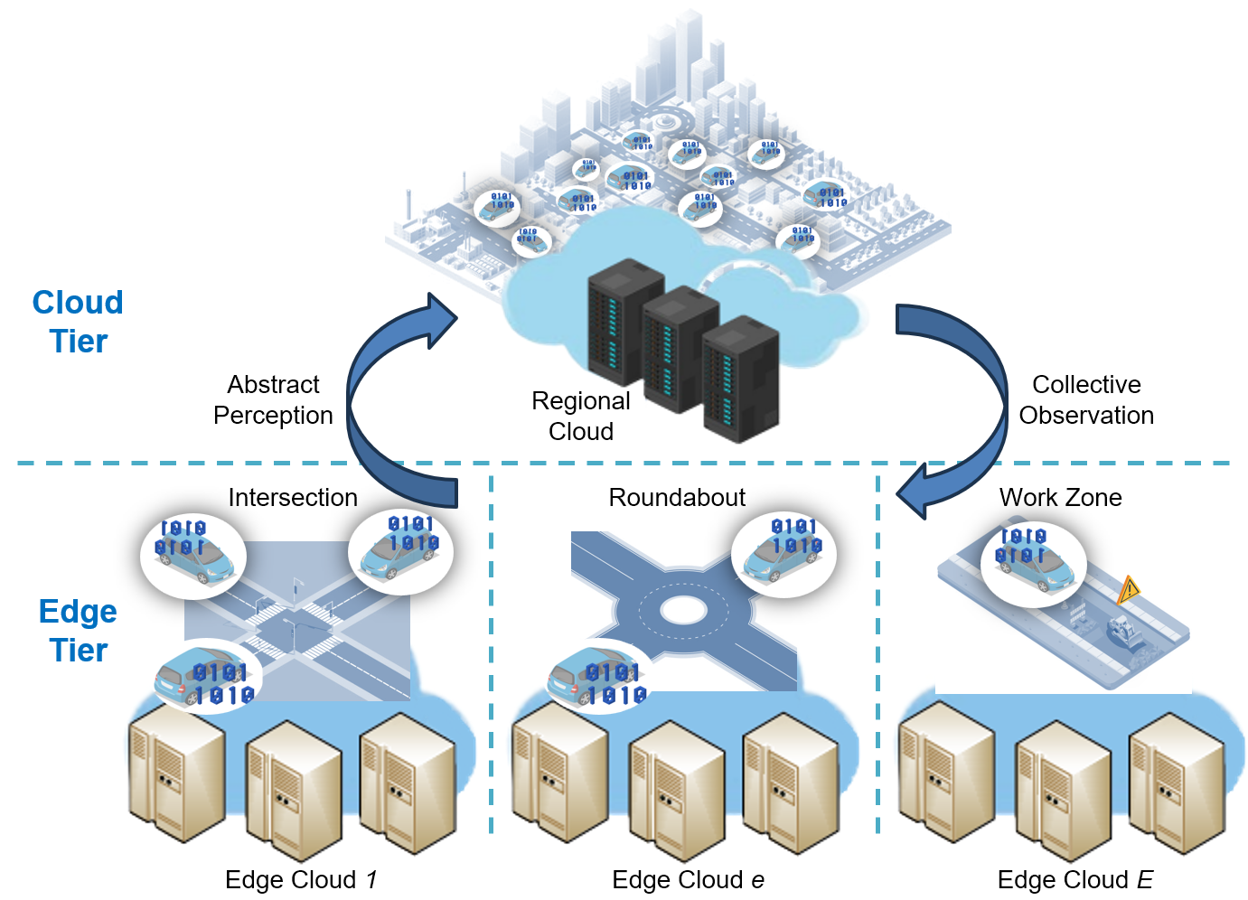} 
  \caption{Proposed two-tier architecture for DT of CAVs.} 
  \label{fig:architecture} 
\end{figure}

This architecture follows a bottom-up approach to create DT of CAVs, as demonstrated in Fig. \ref{fig:architecture}. The tier 1 or the edge tier collects data from CAVs in the scenario and creates a very fine-grained digital replica in ESs. However, in edge tier, only a set of DT is maintained, which does not reflect the whole scenario or collective view. Tier 2 or cloud tier on the other hand collects necessary information of DT instances from the edge tier nodes and creates a collective yet abstract view. The goal of this segregation is to encapsulate the data while exposing only necessary information to applications with different service requirements. For example, time-sensitive applications such as traffic safety programs, requiring highly granular data with very low latency, can interface and operate on the edge tier. On the other hand, the applications which have flexible delay requirement and only need a collective yet coarse-grained data, such as traffic efficiency improvement programs, can operate on the cloud tier. These two tiers are modeled below:

\textbf{\textit{1) Edge Tier}}: 
The nodes of this tier have three main functions: (1) providing edge computing capability to CAVs; (2) managing communication with CAVs through various network functions; and (3) allocating radio resources to APs to ensure quality of service and fairness. At time $t$, a DT $dt_v$ of CAV $v \in \mathcal{V}_r$ is maintained through edge computing of the node $e \in \mathcal{E}$ until the corresponding CAV either stops communicating or moves beyond the communication range. The communication is achieved through communication via APs connected to $e$. A set of $dt_v$ maintained by $e$ can be defined as $\mathcal{D}_e = \{1,2,\ldots,N\}$, where $N$ is the number of $dt_v$ at time $t$. Each node maintains only one DT for each CAV, formally defined as, $\forall v \in \mathcal{V}, \exists! dt_v \in \mathcal{D}_e \text{ such that } DT(v) = dt_v$. However, due to heavy size of the sensor data and the computational expense of DT synchronization, these nodes collaboratively maintain the DTs. Which means the a CAV connected through an AP can have a DT maintained by any of the ESs within the edge network. 

\textbf{\textit{2) Cloud Tier}}:
A set of high performance computing node (CS) denoted as $\mathcal{C} = \{1,2,\ldots,C\}$ are dispersed in different regions. In a region, $\forall e \in \mathcal{E}$ are connected to a $c \in C$ in the cloud tier via a core network. A $c$ perceives refined and required information of $\forall{dt_v}$ generated and maintained at each $e$. As presented in Fig. \ref{fig:architecture}, compared to edge tier which maintains DT for different small scale scenarios such as intersection, roundabout and work zone, the cloud tier creates DT for a city scale. At the same time information for network orchestration such as the available resources of ESs, network topology, link latency, and orchestration policies are also collected at this tier. The CS updates the applications on ESs, which in turn utilize the DTs to provide services. Simultaneously, this tier shares collective information such as available computing resources, with an $e$ on demand during the scheduling process.
\section{Problem Formulation}
\subsection{DT Synchronization Model}
We model the synchronization process of DT by analyzing the delay to update a DT from the data transmitted by it's physical twin, as outlined in \cite{DigitalTwin_sensing}. Synchronization requires CAVs to transmit their data to designated ESs. This data, comprising both $v$'s status and sensing information, is expressed as:
\begin{equation}
    D^t_v = D^{st}_v(t) + D^{se}_v(t)
    \label{eq:eq1}
\end{equation}
where $D^{st}_v(t)$ denotes the status data and $D^{se}_v(t)$ represents the sensing data of vehicle $v$ at time $t$. The communication between vehicles and APs occurs over a wireless medium. The maximum uplink data rate between vehicle $v$ and AP $r$ is given by:
\begin{equation}
    R^{up}_{v,r} = B \log_2\left(1+\frac{p_v h_{vr} \Delta_{v,r}^{-\alpha}}{N}\right)
    \label{eq:eq2}
\end{equation}
where $B$ denotes the channel bandwidth, $p_v h_{vr}$ is the product of the transmission power and channel gain, $\Delta_{v,r}^{-\alpha}$ captures the path loss, $\Delta_{v,r}$ is the distance between vehicle $v$ and AP $r$, and $\alpha$ is the path loss exponent. Assuming that vehicle $v$ uploads data of size $D^t_v$ to AP $r$ during time slot $t$, the corresponding wireless transmission delay is:
\begin{equation}
    T_{v,r} = \frac{D^t_v}{R^{up}_{v,r}}
    \label{eq:wireless_delay}
\end{equation}
Simultaneously, the wired transmission delay between AP $r$ and ES $e \in \mathcal{E}$ must also be considered. This delay depends on both the data size and the number of hops in between, and is expressed as:
\begin{equation}
    T_{r,e} = \lambda \Delta_{r,e} D^t_v 
    \label{eq:wired_delay}
\end{equation}
where $\lambda$ denotes the processing and transmission delay per unit data for every hop within the network, and $\Delta_{r,e}$ represents the number of hops between AP $r$ and ES $e$. Upon arrival at the $e$, the data $D^t_v$ is processed to update and synchronize the digital twin $dt_v$. The processing time to synchronize $dt_v$ is:
\begin{equation}
    T_{dt_v} = \max\left\{\frac{f(D^t_v)}{CPU_e},\frac{g(D^t_v)}{GPU_e},\frac{h(D^t_v)}{Store_e}\right\}
    \label{eq:processing_delay}
\end{equation}
where $f(D^t_v)$, $g(D^t_v)$, and $h(D^t_v)$ represent the CPU, GPU, and storage operations required for processing $D^t_v$, respectively, and $CPU_e$, $GPU_e$, and $Store_e$ denote the execution capabilities of ES $e$ per unit time for CPU, GPU, and storage operations. It is assumed that these operations are executed concurrently, and the longest running operation determines $T_{DT_v}$. Therefore, when $dt_v$ is hosted on $e$, the total DT synchronization latency is given by:
\begin{equation}
    T^{syn}_{dt_v} = T_{v,r} + T_{r,e} + T_{dt_v}
    \label{eq:total_delay}
\end{equation}
The performance of this process is measured based on the synchronization error, which indicates whether the DT update interval is too long for effective synchronization. The DT update interval is defined as the time between two consecutive updates of the DT, and it must satisfy the following condition:
\begin{equation}
    T^{syn}_{dt_v} \leq T_{th}
    \label{eq:delay_threshold}
\end{equation}
where $T_{th}$ represents a predefined threshold. If this threshold is exceeded, the DT becomes misaligned with its physical counterpart, resulting in synchronization error.

\subsection{Collaborative Computation Model}

In this section, we introduce the collaborative computing resource provisioning model and define the overall utilization of multidimensional resources to assess the performance of the proposed algorithm.

In the proposed framework, which integrates communication and computation, an $e$ is aware of the computational capability and available resources of $\forall e \in \mathcal{E}$ by making query to the connected cloud server $c$.
We also assumes that CAVs leverage their multi-modal sensing capabilities to generate $D^t_v$. Once a $v$ establishes a connection with an $r$, a tuple is placed in a queue $Q$, defined by: 
\begin{equation}
Q = [v_1, v_2, \dots, v_n]
\label{eq:queue}
\end{equation}
Each element stored in $Q$ is a data from $v$ to be formulated into a computing task. The tuple contains $D^t_v$ along with the service demands of $v$ to create and synchronize $dt_v$. A computing task can be described according to service demands and defined based on major resource requirements \cite{zhang2024two}. Since the service demands for synchronizing a DT using $D^t_v$ are based on the CPU, GPU and storage resource requirements, a DT task generated from $v$ can be defined as: 
\begin{equation}
    Task_{dt_v} = \{f(D^t_v), g(D^t_v), h(D^t_v), D^t_v, src, dst\}
    \label{eq:def_of_task}
\end{equation}
where each task includes the required GPU, CPU, and storage operation, the size of the data to be processed, the source address ($src$) of the CAV from which the data is received, and the destination address ($dst$) of the $e$ where the task should be processed. However, as the ESs simultaneously perform multiple roles, maintaining scalability by accommodating a growing number of tasks from CAVs within the same computing cluster becomes increasingly challenging. Hence, we propose collaborative computation among ESs and formulate a computing resource aware dynamic task provisioning problem. This problem leads to the selection of $dst$ based on a binary  decision variable $x(dt_v,e)$, defined as:
\begin{equation}
x(dt_v,e) = 
\begin{cases}
1, & \text{if } e \text{ is used to process task of }dt_v\\
0, & \text{otherwise}
\end{cases}
\label{eq:resource_decision}
\end{equation}
To ensure scalability and uninterrupted operation while preventing the over-utilization of resources, we establish the following set of resource constraints:
\begin{equation}
    CPU^{use}_e = CPU^{r}_e + \sum_{d \in \mathcal{D}_e} cpu_d \leq CPU_e
    \label{eq:cpu_constraint}
\end{equation}
\begin{equation}
    GPU^{use}_e = GPU^{r}_e + \sum_{d \in \mathcal{D}_e} gpu_d \leq GPU_e
    \label{eq:gpu_constraint}
\end{equation}
\begin{equation}
    Store^{use}_e = Store^{r}_e + \sum_{d \in \mathcal{D}_e} store_d \leq Store_e
    \label{eq:storage_constraint}
\end{equation}
\begin{equation}
    x(dt_v,e) = 1, \quad \forall dt_v
    \label{eq:mapping_constraint}
\end{equation}
Here, $CPU^{r}_e$, $GPU^{r}_e$, and $Store^{r}_e$ denote the resources reserved by edge server $e$ for background processes, network functions, and radio resource management. The variables $cpu_d$, $gpu_d$, and $store_d$ represent the computing resources required to process the DT task $d$ on edge server $e$. To evaluate the solution to the resource provisioning problem, we use the metric defined in equation (\ref{eq:resource_utilization}), following the approach in \cite{zhang2024two}, across all computing domains in $\mathcal{E}$.

\begin{equation}
    \sum_{e \in \mathcal{E}}(UR_e) = \frac{\sum_{e \in \mathcal{E}} (UR^{gpu}_e + UR^{cpu}_e + UR^{store}_e)}{3 |\mathcal{E}|}.
\label{eq:resource_utilization}   
\end{equation}

\begin{figure}[t]
  \centering
    \includegraphics[width=3.4in]{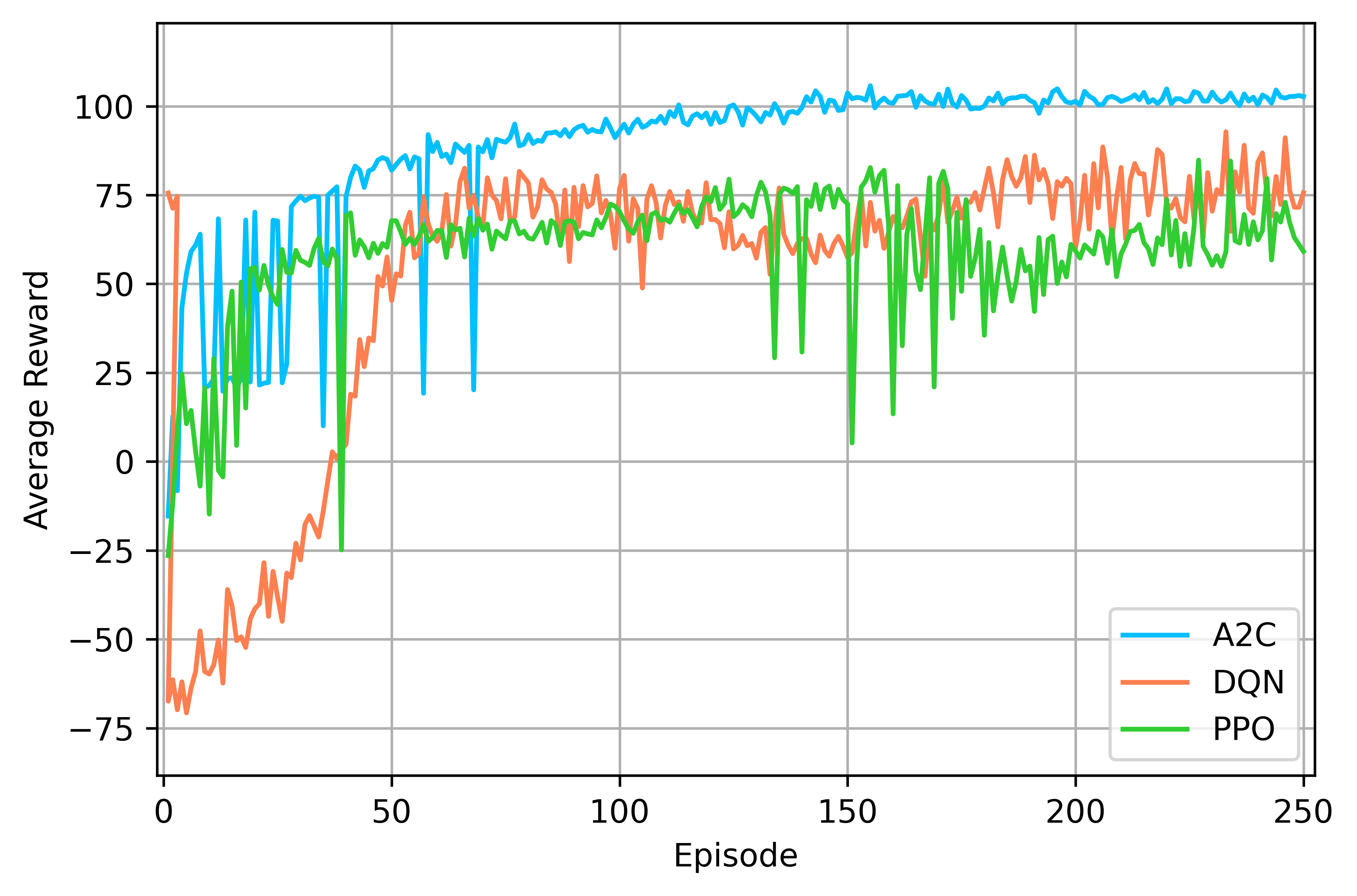} 
  \caption{Episode-wise average reward during training.} 
  \label{fig:reward} 
\end{figure}

\subsection{Formulation of Optimization Problem}

This study aims to improve overall system performance in a distributed DT environment by optimizing computing resource allocation and minimizing synchronization delays between physical entities and their virtual counterparts. From the preceding analysis, the challenge of provisioning computing resources efficiently is modeled as a multi-objective optimization problem:
\begin{equation}
\begin{aligned}
\max_{v \in \mathcal{V}_r} \left( \sum_{e \in \mathcal{E}} UR_e - T^{syn}_{dt_v} \right) \\
\text{s.t. (\ref{eq:delay_threshold}), (\ref{eq:cpu_constraint}) - (\ref{eq:mapping_constraint})}.
\label{eq:objective}
\end{aligned}
\end{equation}
Here, $UR_e$ denotes the utilization rate of computing resource $e \in \mathcal{E}$, and $T^{syn}_{dt_v}$ represents the synchronization latency for DT instance of $v$. The objective function captures the trade-off between resource utilization and synchronization performance, which are critical for ensuring real-time responsiveness and efficient operation of digital twins in heterogeneous edge-cloud environments. The problem is further complicated by the stochastic arrival of tasks, variable resource demands, and the dynamic availability of heterogeneous resources (e.g., CPU, GPU, and storage), which shows both spatial and temporal variability. Moreover, resource provisioning decisions are interdependent across sub-systems, leading to a coupling of sub-problems that renders traditional optimization methods computationally infeasible in real-time scenarios.
To address these challenges, we propose a Deep Reinforcement Learning (DRL)-based resource provisioning framework. This approach leverages the ability of DRL agents to learn optimal policies from interaction with the environment, enabling adaptive and scalable decision-making under uncertainty.
\section{Intelligent Edge-Resource Provisioning Algorithm}
The proposed resource provisioning algorithm operates on $Q$ containing unassigned future tasks at each decision making step. Once an provisioning decision is made for a $Task^{DT}_v$, it remains unchanged until the destination $e$ is relieved of its assignment. Since the assignment of the next computing task transmission depends solely on the current available computing resources and the associated delay state, without being influenced by the transmission history of prior tasks—the problem naturally conforms to the structure of a Markov Decision Process (MDP). 
Specifically, the task manager is modeled as an agent interacting with the system environment, where the state space encodes current resource usage, task demands, and synchronization statuses; the action space corresponds to possible provisioning strategies, and the reward function is designed to balance high resource utilization with low synchronization delay. By continuously learning from feedback, the DRL agent can approximate near-optimal provisioning policies even in highly dynamic and complex environments.

\begin{figure}[t]
  \centering
    \includegraphics[width=3.4in]{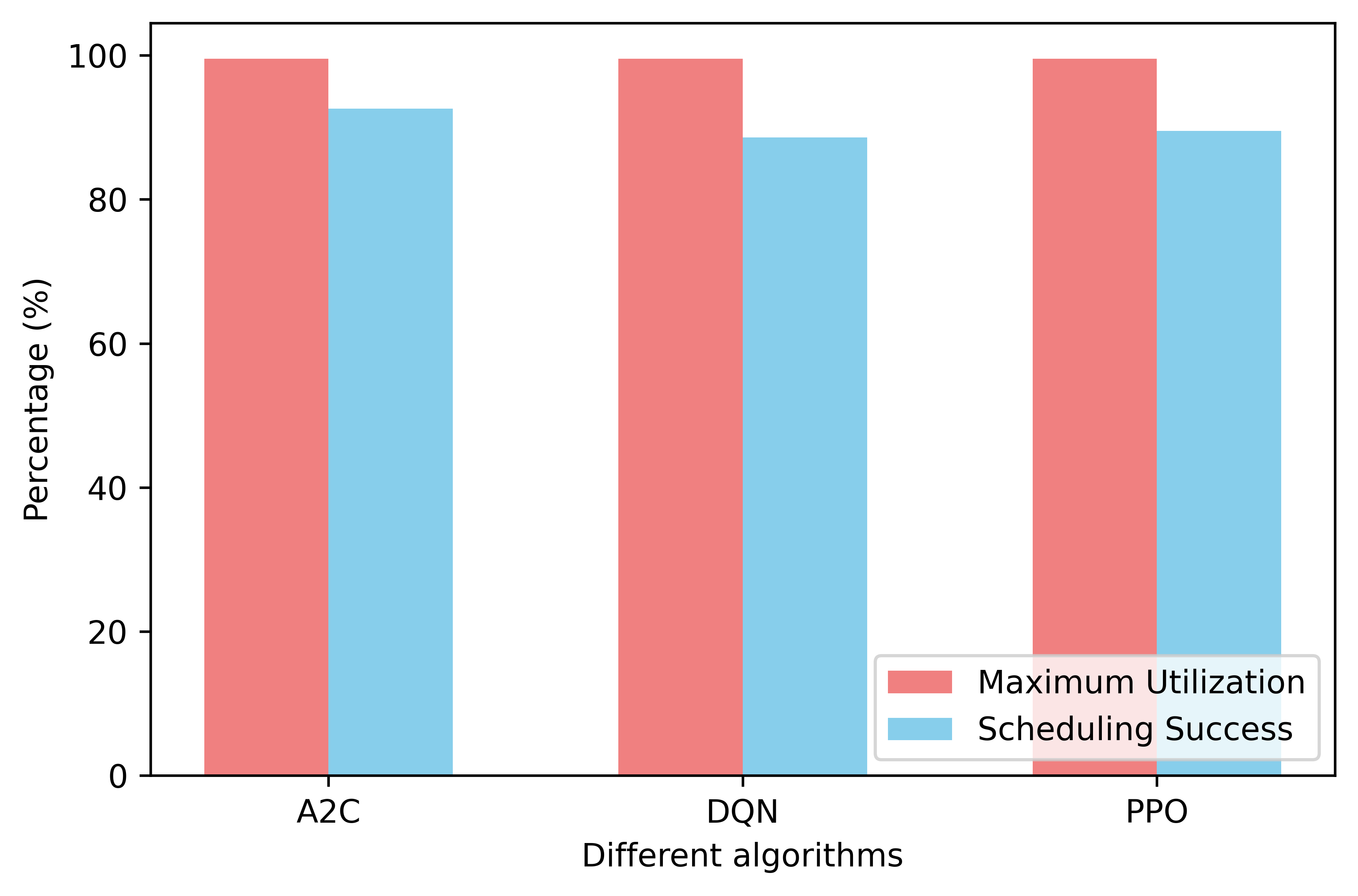} 
  \caption{Comparison of two performance metrics across different scheduling algorithms.} 
  \label{fig:algo_comparison} 
\end{figure}

\begin{figure*}[t]
  \centering
    \includegraphics[width=7.0in]{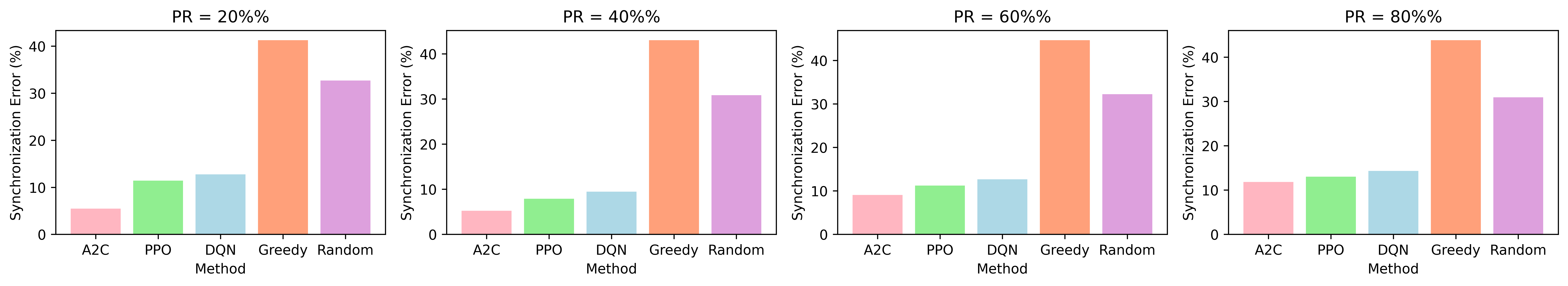} 
  \caption{Comparison of synchronization error for three control strategies across varying CAV penetration ratios (20\%–80\%).} 
  \label{fig:per_vs_sync} 
\end{figure*} 

\subsection{State}
At each episode, the agent deployed at ES $e$ collects transmission delay information for a CAV $v \in \mathcal{V}_r$ by calculating $T_{v,r}$, where $r \in \mathcal{R}_e$, and retrieves $T_{r,e}$ for all $e \in \mathcal{E}$ from the CS. In addition, the agent obtains information about the capacity and available resources of the computing domains from the CS. The system state denoted by $S$, is defined as:
\begin{equation}
\begin{aligned}
S = \left(\Theta^T_v, \Theta^C_e, \Theta^U_e, \Theta^C_v \right), \quad \quad \quad \quad \\ v = v_1, \; Q \leftarrow [v_2, \dots, v_n], \forall e \in \mathcal{E}, \ t \in \mathcal{T}
\label{eq:state}
\end{aligned}
\end{equation}
where $\Theta^T_v = T_{v,r} + T_{r,e}$ represents the total transmission delay for $ \forall e \in \mathcal{E}$, $\Theta^C_e = [\text{CPU}_e, \text{GPU}_e, \text{Store}_e]$ describes the computing capacity of $ \forall e \in \mathcal{E}$, $\Theta^U_e = [\textit{CPU}^e_t, \textit{GPU}^e_t, \textit{Store}^e_t]$ indicates the utilized computing resources of $ \forall e \in \mathcal{E}$ at step $t$, and $\Theta^C_v = [f(D^t_v), g(D^t_v), h(D^t_v)]$ denotes the computing requirements for updating $d^t_v$.

\subsection{Action} 

Based on the observed state, the agent makes real-time task assignment decisions by selecting a computing node $e$ to process the DT $dt_v$ of CAV $v$. The action space $A$ is defined as:
\begin{equation}
A = \mathcal{E} = \{1, 2, \ldots, E\}, \quad a_i \in A, |\{a_i\}| = 1
\end{equation}
where $|\{a_i\}| = 1$ ensures that at step $i$, only one node $e \in \mathcal{E}$ is selected to execute each task, in accordance with the constraint specified in equation~(\ref{eq:mapping_constraint}).

\subsection{Reward} 
Once the agent takes action $a_i$, it receives a reward $r_i$ to evaluate the effectiveness of its decision and to simultaneously maximize computing resource utilization and minimizing the synchronization delay. The episodic reward $R$ for the learning-based algorithm is defined as:
\begin{equation}
\begin{aligned}
  R &= \frac{\sum^{M}_{i=1} r_i}{M}, \\
  r_i &= \delta 
        + \epsilon \sum_{e \in \mathcal{E}} UR_e
        - \zeta\,T^{\mathrm{syn}}_{dt_v}, \\
  \delta &=
  \begin{cases}
    \delta_1 > 0,
      &\text{if }
      \begin{aligned}[t]
        CPU^{\mathrm{use}}_e &\le CPU_e,\\
        GPU^{\mathrm{use}}_e &\le GPU_e,\\
        Store^{\mathrm{use}}_e &\le Store_e
      \end{aligned}
    \\[1ex]
    \delta_2 < 0, &\text{otherwise.}
  \end{cases} 
  \label{eq:reward}
\end{aligned}
\end{equation}
In equation~(\ref{eq:reward}), $\delta$ represents a high-magnitude value used to reward or penalize the agent based on compliance with the constraints defined in equations~(\ref{eq:cpu_constraint}), (\ref{eq:gpu_constraint}), and (\ref{eq:storage_constraint}). The terms $\epsilon$ and $\zeta$ serve as weighting coefficients to balance the individual components of the objective.

Within the MDP formulation introduced above, the edge-side agent seeks a \textit{task-provisioning policy} that decides, at every decision epoch, which computing node $e \in \mathcal{E}$ should process the pending DT update $dt_v$. Formally, the optimization problem is defined as:
\begin{equation}
\begin{aligned}
\max_{\;\pi \in \Pi}\;
\mathbb{E}\left[
      \sum_{k=0}^{\infty} \gamma^{k}\, r_{i+k}
      \;\middle|\;
      s_i, a_i;\,\pi
\right],
 \\
  \text{s.t. (\ref{eq:delay_threshold}), (\ref{eq:cpu_constraint}) - (\ref{eq:mapping_constraint})}, (\ref{eq:state}), (\ref{eq:reward}).
\end{aligned}
\label{eq:policy}
\end{equation}
Here, the expression $\max_{\pi \in \Pi}$ indicates that the goal is to find the optimal policy $\pi^*$ from the set of all possible policies $\Pi$. The expectation term $\mathbb{E}[\cdot \mid s_i, a_i; \pi]$ captures the expected cumulative reward starting from a given state $s_i$ and action $a_i$, assuming the agent follows policy $\pi$ thereafter. The summation $\sum_{k=0}^{\infty} \gamma^k r_{i+k}$ represents the discounted return, which is the total future reward accumulated starting from step $i$, with each future reward $r_{i+k}$ scaled by the discount factor $\gamma^k$. The term $r_{i+k}$ refers to the reward received at time $i + k$, and $\gamma^k$ applies exponential discounting, where $\gamma \in [0, 1)$, reducing the influence of rewards further in the future. The learned policy $\pi$ maps each observed state $S$, comprising transmission delays, available and utilized CPU/GPU/storage resources, and the requirements of the current task to an action $a_t = e$, thereby dynamically assigning DT updates to appropriate edge computing nodes with the aim of maximizing the long-term expected return.


\section{Performance Evaluation}
\subsection{Experimental Setup}
To evaluate the decision-making capability of the resource provisioning algorithm and the scalability of the proposed framework, we develop a SUMO traffic simulation using real traffic volume data extracted from drone footage of road traffic near NC State University, Raleigh, USA \cite{DT_roundabout}. For each episode, we simulate 1986 vehicles for different routes, sample CAVs at different market-penetration ratio (PR), and generate data for DT tasks, each with its own computational resource requirements. Additionally, we randomly instantiate \(E = 4\) edge nodes per episode, each with distinct computing capabilities and a varying number of network hops. For each ES node, an AP is placed at a random location, which connects to the CAVs within the communication range. The value of $T_{th}$ is set to $25\,\text{ms}$ for measuring synchronization errors, while instances where the algorithm attempts to over-utilize resources are used to assess task scheduling success percentage. In addition to comparing the performance of the employed A2C, DQN, and PPO algorithms for training the agent, we also compare the performance with both a random approach and a greedy approach. The random approach selects an ES node at random for computation, whereas the greedy method aims to maximize resource utilization by selecting nodes in a specific order, provided their computing capacity is not fully utilized.

\subsection{Simulation Results}
As illustrated in Fig.~\ref{fig:reward}, the episode-wise rewards observed during training demonstrate the agent's performance improvement over successive episodes, with the A2C algorithm achieving higher rewards compared to the other two algorithms.
In contrast, Fig.~\ref{fig:algo_comparison} provides a comparative analysis of the maximum resource utilization of ES nodes and the task scheduling success rates achieved by three algorithms. While all the algorithms achieve high maximum utilization, A2C shows superior scheduling success relative to the other two. The synchronization error increases as the number of tasks grows with rising PR in the simulation as depicted in Fig. \ref{fig:per_vs_sync}. The greedy and random methods show consistently high synchronization errors across all PR values, as they do not consider synchronization delay minimization. In comparison, the DRL based algorithms perform significantly better. Among them, A2C achieves the lowest synchronization error, with values dropping to just 5\% at a PR of 20\%, and remaining as low as 12\% even when 80\% of vehicles generate DT tasks.

\section*{Conclusion}
Despite the growing interest in DT for ITS, their uninterrupted operation under real-time, data-intensive conditions remains an open challenge. This work presents a distributed computing architecture that brings together DTs and MEC within a software-defined vehicular networking context. Through the development of a collaborative task computation algorithm and an DRL based decision-making agent, the framework effectively addresses key limitations in synchronization latency and resource allocation. Simulation results show that DRL maintains a balanced and consistently high performance across different matrices, demonstrating its effectiveness in reliable computing resource scheduling. Further evaluations based on realistic CAV traffic simulations, demonstrate that the system not only maintains DT synchronization with errors as low as 5\% but also maximizes edge resource utilization.

\section*{Acknowledgement}
This material is partially based upon work supported by the National Science Foundation (NSF) under Grant CNS-2210344, and the NCDOT under Award TCE2020-03. The contents do not necessarily reflect the official views or policies of the North Carolina Department of Transportation (NCDOT). This paper does not constitute a standard, specification, or regulation.

\begingroup
\footnotesize
\bibliographystyle{ieeetr}
\bibliography{cite}
\endgroup
\end{document}